\documentclass{aa}  
\usepackage{graphicx}
\usepackage{txfonts}
\usepackage[colorlinks=true,linkcolor=blue,citecolor=blue,urlcolor=black]{hyperref}
\usepackage{natbib}
\bibpunct{(}{)}{;}{a}{}{,}

\begin{document} 
   \title{Study of the sub-AU disk of the Herbig B[e] star HD~85567
     with near-infrared interferometry \thanks{Based on observations
       made with ESO telescopes at the La Silla Paranal Observatory
       under program IDs 080.C-0541(C), 082.C-0893(A), 084.C-0848(B)}
   } \titlerunning{The sub-AU disk of HD 85567}

   \author{
     J. Vural \inst{1} \thanks{Member of the International Max
       Planck Research School (IMPRS) for Astronomy and Astrophysics
       at the Universities of Bonn and Cologne}, 
     S. Kraus \inst{2},
     A. Kreplin \inst{1},
     G. Weigelt \inst{1}, 
     E. Fossat \inst{3},
     F. Massi \inst{4},
     K. Perraut \inst{5},
     F. Vakili \inst{3}
   }

   \institute{
     Max-Planck-Institut f\"ur Radioastronomie, Auf dem H\"ugel 69,
     53121 Bonn, Germany
     \and University of Exeter, School of Physics, Physics Building,
     Stocker Road, Exeter, EX4 4QL, UK
     \and Laboratoire Lagrange, UMR7293, Universit\'e de Nice
     Sophia-Antipolis, CNRS, Observatoire de la C\^ote d'Azur, 06300
     Nice, France 
     \and INAF - Osservatorio Astrofisico di Arcetri, Largo E. Fermi,
     5, 50125 Firenze, Italy 
     \and UJF-Grenoble 1 / CNRS-INSU, Institut de Plan{\'e}tologie et
     d'Astrophysique de Grenoble (IPAG) UMR 5274, Grenoble, F-38041,
     France
   }

   \date{Received ...; accepted 10 July 2014}

% \abstract{}{}{}{}{} 
% 5 {} token are mandatory
 
  \abstract
  {The structure of the inner disk of Herbig Be stars is not well
    understood. The continuum disks of several Herbig Be stars have
    inner radii that are smaller than predicted by models of
    irradiated disks with optically thin holes. }
  {We study the size of the inner disk of the Herbig B[e] star
    HD~85567 and compare the model radii with the radius suggested
    by the size-luminosity relation.}
  {The object was observed with the AMBER instrument of the Very Large
    Telescope Interferometer. We obtained {\textit K}-band
    visibilities and closure phases. These measurements are
    interpreted with geometric models and temperature-gradient
    models. }
  {Using several types of geometric star-disk and star-disk-halo
    models, we derived inner ring-fit radii in the {\textit K} band
    that are in the range of 0.8--1.6~AU. Additional
    temperature-gradient modeling resulted in an extended disk with an
    inner radius of $0.67^{+0.51}_{-0.21}$~AU, a high inner
    temperature of $2200^{+750}_{-350}$~K, and a disk inclination of
    $53^{+15}_{-11}$\degr.}
  {The derived geometric ring-fit radii are approximately 3--5 times
    smaller than that predicted by the size-luminosity relation. The
    small geometric and temperature-gradient radii suggest optically
    thick gaseous material that absorbs stellar radiation inside the
    dust disk.}


   \maketitle
%
%________________________________________________________________
\section{Introduction}
Infrared interferometric observations suggest that the circumstellar
environment of Herbig Be (HBe) and Herbig Ae (HAe) stars are
significantly different. In contrast to the lower-mass HAe stars, the
{\textit K}-band continuum radii of several HBe stars are
significantly smaller than predicted by the size-luminosity relation
\citep{2005monmil,2004eislan,2008krapre,2011weigri,2012krekra}. These
smaller radii can be explained by the presence of an optically thick
gas inside the dust disk. This optically thick gas can absorb the
stellar ultraviolet (UV) radiation and allows dust to exist closer to
the star \citep{2002monmil}.

The object \object{HD~85567} (V596~Car, Hen~3-331) is a B-type star at
a distance of $1.5 \pm 0.5$~kpc (see stellar parameters in
Table~\ref{tabpro}). The evolutionary status of HD~85567 is not yet
well-established. \citet{2001mirlev} suggested the object to be a
main-sequence B[e] star. Other studies reported that HD~85567 is a
young stellar object
\citep{1998lamzic,1998malbog,2012verwat,2013whewei}.
\citet{2001mirlev} proposed the existence of a close binary companion
that can interact with the circumstellar disk of HD~85567. Binarity is
believed to be a key property in B[e] stars
\citep{2007mir}. \citet{2006baioud} detected a binary companion with a
separation of $\gtrsim$$\,500$~mas.

We use our near-infrared (NIR) interferometric observations to
investigate the inner disk structure of HD~85567.  The paper is
organized as follows. We describe the observations and the data
reduction in Sect.~\ref{kapobs}. The modeling is presented in
Sect.~\ref{kapmod} and the results are discussed in
Sect.~\ref{kapdis}.

\begin{table}[t]
\caption{Stellar parameters of HD~85567.}     
\label{tabpro}   
\centering                       
\begin{tabular}{rcl}  
\hline\hline
Parameter               & Value                         \\ \hline
spectral type           & B2V             \tablefootmark{~(a)} \\
distance [kpc]          & $1.5 \pm 0.5$   \tablefootmark{~(a)} \\
$M_* [M_\odot]$          & $12 \pm 2$      \tablefootmark{~(b)} \\
$\log(T_* [\mathrm K])$ & $4.32 \pm 0.08$ \tablefootmark{~(b)} \\
$\log(L_* [L_\odot])$    & $4.17 \pm 0.16$ \tablefootmark{~(b)} \\
$R_* [R_\odot]$          & $9 \pm 2$       \tablefootmark{~(b)} \\
$A_\mathrm{V}$           & $1.1 \pm 0.1$   \tablefootmark{~(b)} \\
$\dot M [10^{-6} M_\odot \mathrm{yr}^{-1}]$  & $6.3^{+4}_{-2}$ \tablefootmark{~(c)}\\
\hline
\end{tabular}
\tablefoot{References: \tablefoottext{a}{\citet{2001mirlev}},
  \tablefoottext{b}{\citet{2012verwat}},
  \tablefoottext{c}{\citet{2013ile}} .}
\end{table}

%__________________________________________________________________
\section{Observation and data reduction} \label{kapobs}
\begin{figure}[t]
  \includegraphics[angle=-90,width=0.4\textwidth]{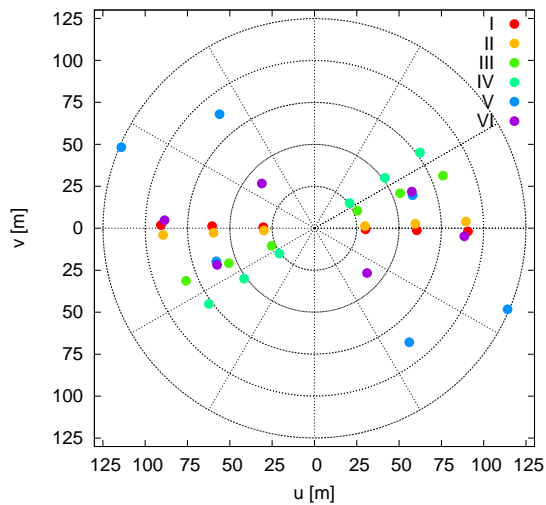}
  \caption{The $uv$ coverage of our AMBER measurements of HD~85567
    (see Table~\ref{tabobs}).}
  \label{figuv}
\end{figure}

The observations of HD~85567 were carried out using the near-infrared
three-beam combiner VLTI/AMBER \citep{2007petmal} in three different
nights. We obtained a total of six measurements in the low spectral
resolution mode ($R=30$: Fig.~\ref{figuv}, Table~\ref{tabobs}). The
data were reduced with \textit{amdlib
  3.0.5}\footnote{\url{http://www.jmmc.fr/data_processing_amber.htm}}
\citep{2007tatmil,2009cheutr}. We applied a signal-to-noise frame
selection (20\% of the highest fringe signal-to-noise ratio (SNR);
\citealt{2007tatmil}) to the raw files of object and calibrator
to obtain an improved visibility calibration. An additional
improvement of the visibility calibration was achieved by equalizing
the optical path difference (OPD) histograms of the object and
calibrator \citep{2012krekra} to account for atmospheric OPD drifts.

\begin{figure}[t]
  \centering
  \includegraphics[width=0.48\textwidth]{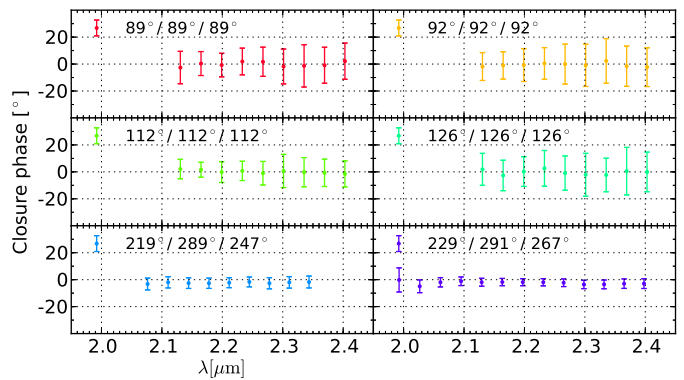}
  \caption{Derived closure phases. The color coding is the same as in
    Fig.~\ref{figuv}.}
  \label{figcp}
\end{figure}

\begin{figure*}[t!]
  \centering
  \includegraphics[width=\textwidth]{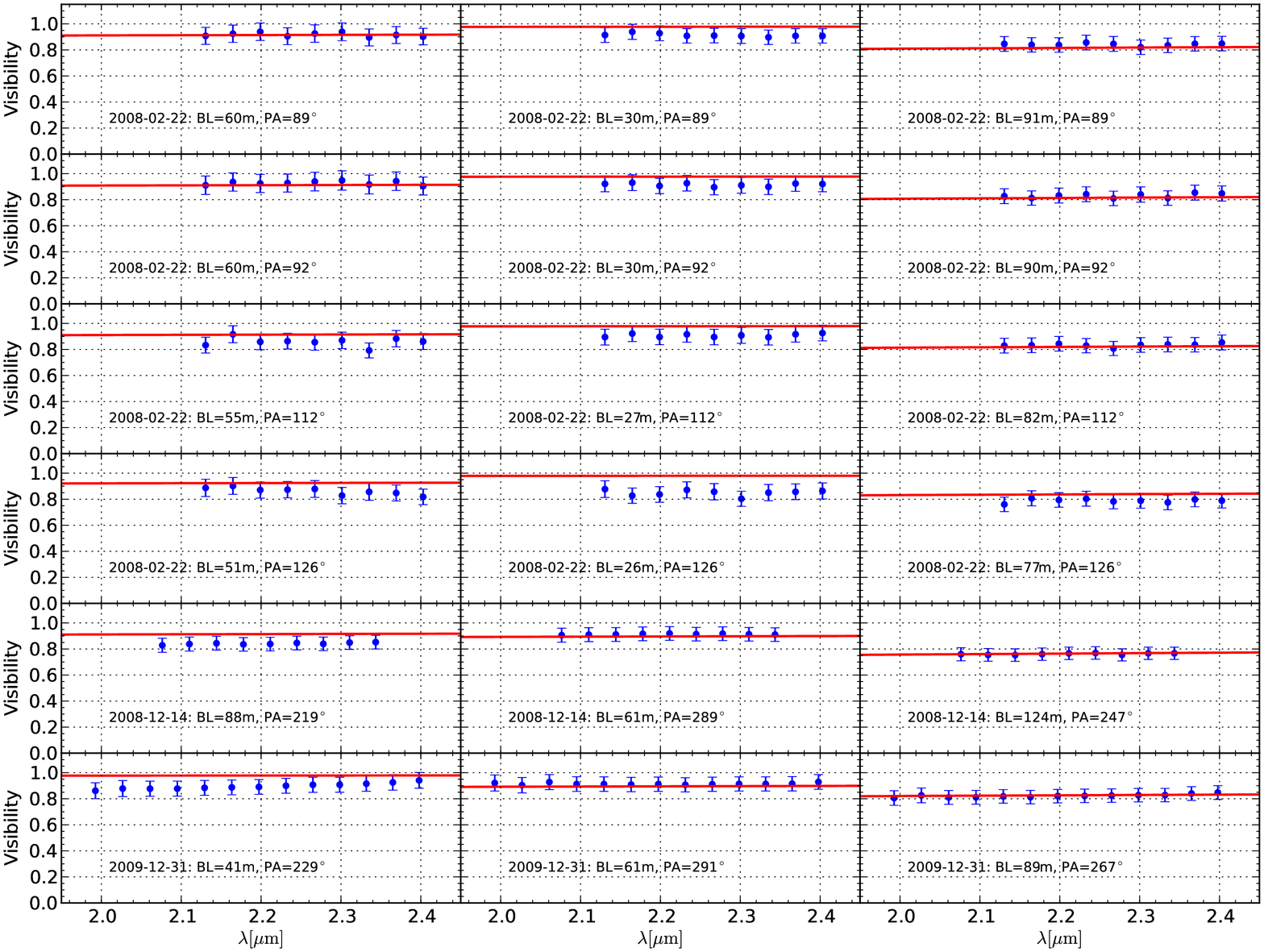}
  \caption{Comparison of the visibilities of the best-fit
    temperature-gradient model (red line; see Table~\ref{tabpar}) with
    the wavelength-dependent AMBER visibilities (blue dots). The
    wavelength range of the visibilities is different for all
    measurements because of different data quality as explained in
    Sect.~\ref{kapobs}.}
  \label{figvis} 
\vspace*{2mm}
  \begin{minipage}{0.55\textwidth}
    \centering
    \includegraphics[width=\textwidth]{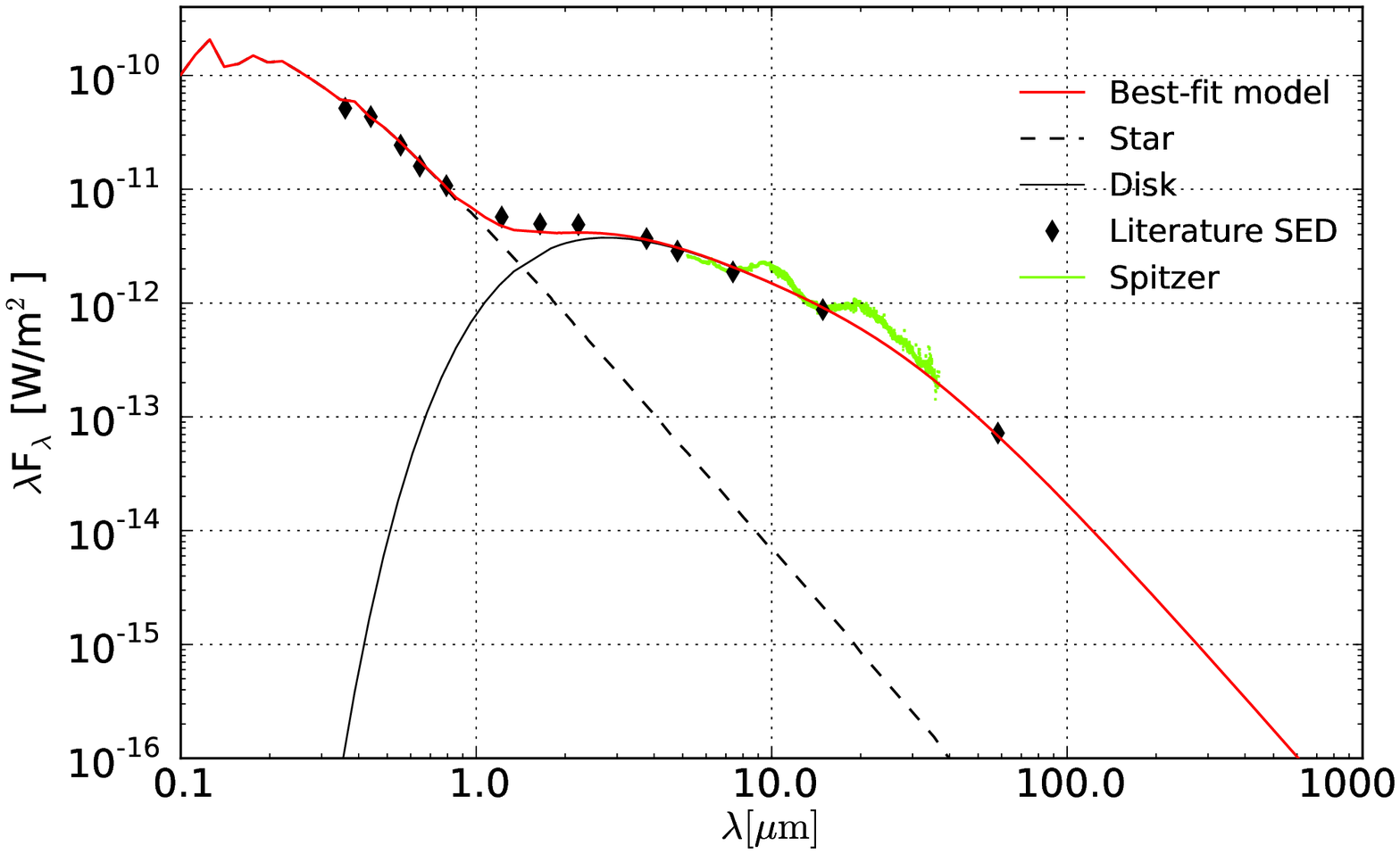}
  \end{minipage}
  \begin{minipage}{0.44\textwidth}
    \centering
    \includegraphics[angle=-90,width=0.8\textwidth]{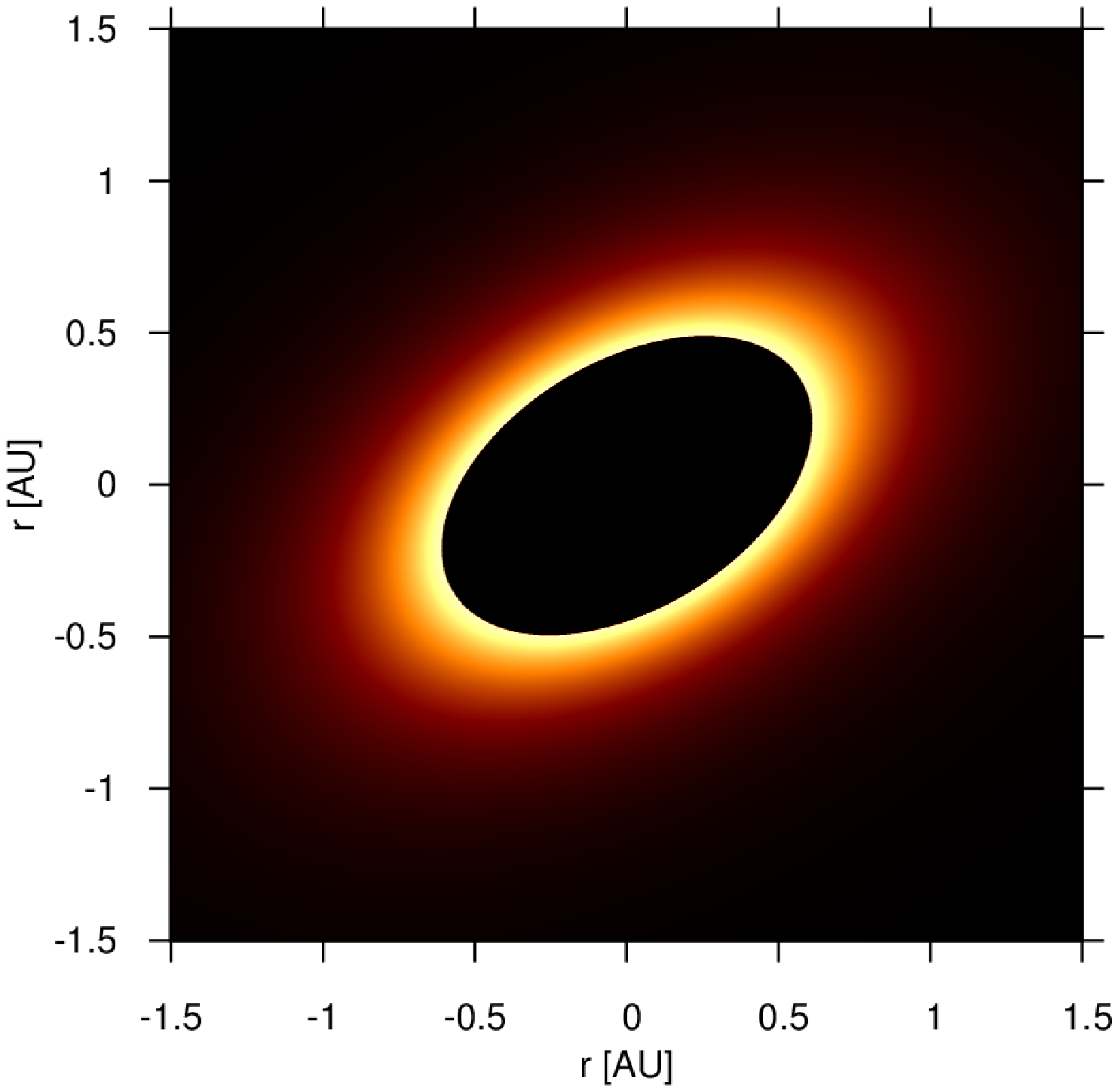}
  \end{minipage}
  \caption{{\it Left:} Observed SED (black diamonds) and SED of the
    best-fit temperature-gradient model (Table~\ref{tabpar}). The
    total SED (red line) consists of the stellar flux (Kurucz model,
    black dashed line) and the flux of the temperature-gradient disk
    (black solid line). {\it Right:} Two-dimensional intensity
    distribution of the best-fit model at $2~\mu$m. The color-scaling
    is linear in arbitrary units. The star is not displayed. }
  \label{figsed} 
\end{figure*}

We derived {\textit K}-band closure phases (Fig.~\ref{figcp}) and
visibilities (Fig.~\ref{figvis}). The closure phases are zero within
the error bars, indicating that the brightness distribution of our
source is centrosymmetric. Except in measurement VI, \textit{amdlib
  3.0.5} did not compute visibilities and closure phases for
wavelengths in the range of approximately 2.0 -- 2.1~$\mu$m because of
low SNR. The SNR of the {\textit H}-band data was also too low to
derive {\textit H}-band visibilities and closure phases.

The spectral energy distribution (SED; see Fig.~\ref{figsed}) was
reconstructed from dereddened values found in the literature
\citep{2012verwat} and data from the Spitzer Space Telescope (IRS,
program ID: 3470).

\begin{table*}[htb]
  \caption{AMBER observation log.}
  \label{tabobs}   
  \centering                       
  \begin{tabular}{rrrrrrrrrr}  
    \hline\hline
    \#  & Night      & Configuration & B$_\mathrm{proj}$ & PA          & Seeing    & DIT  & Calibrator & Calibrator diameter \\
        &            &               &  [m]             & [\degr]     & [\arcsec] & [ms] &            & [mas]    \\ \hline
    I   & 2008-02-22 &  A0-D0-H0     & 60/91/30         & 89/89/89    & 0.65      & 21   & HD~50281    & $1.0   \pm 0.1  $ \tablefootmark{~(a)} \\
    II  & 2008-02-22 &  A0-D0-H0     & 60/91/30         & 92/92/92    & 0.72      & 100  & HD~85313    & $0.451 \pm 0.045$ \tablefootmark{~(b)} \\
    III & 2008-02-22 &  A0-D0-H0     & 55/82/27         & 112/112/112 & 0.69      & 100  & HD~85313    & $0.451 \pm 0.045$ \tablefootmark{~(b)} \\
    IV  & 2008-02-22 &  A0-D0-H0     & 51/77/26         & 126/126/126 & 0.57      & 100  & HD~85313    & $0.451 \pm 0.045$ \tablefootmark{~(b)} \\
    V   & 2008-12-14 &  U1-U3-U4     & 88/61/124        & 219/289/247 & -         & 26   & HD~50277    & $0.356 \pm 0.036$ \tablefootmark{~(b)} \\
    VI  & 2009-12-31 &  U2-U3-U4     & 41/62/89         & 229/291/267 & 1.1       & 21   & HD~85313    & $0.451 \pm 0.045$ \tablefootmark{~(b)} \\
    \hline
  \end{tabular}
  \tablefoot{The resulting $uv$ coverage is shown in
    Fig.~\ref{figuv}. B$_\mathrm{proj}$ describes the projected
    baseline length, PA the position angles of the baselines, and DIT
    the detector integration time. References:
    \tablefoottext{a}{\citet{2001paspas}},
    \tablefoottext{b}{\citet{2010lafmel}}. We assumed the calibrator
    diameter error to be 10\%.}
\end{table*}

%__________________________________________________________________
\section{Modeling} \label{kapmod}
\subsection{Geometric modeling} \label{kapgeo}
The NIR emission of Herbig stars is believed to originate mainly in a
ring-like region at the inner edge of the disk, which is often
associated with a puffed-up inner rim
\citep{2001natpru,2001duldom}. Thus, the size of the NIR emission
region is often approximated with a geometric ring model, and the
ring-fit radius can be compared to theoretical predictions
\citep{2001milsch,2002monmil,2003eislan,2010dulmon}.

Here, we apply several different ring models, all containing an
unresolved stellar component and a ring with a width of 20\% of the
inner ring radius \citep{2005monmil}. We also investigate inclination
effects (i.e., elongated rings) and the influence of an extended halo
(see Table~\ref{tabgeo}).

The variables of the two-dimensional total visibility
$|V|=|V(PA,\nu_\mathrm{uv})|$ are the position angle (PA) of the
measurement and the spatial frequency $\nu_\mathrm{uv} =
B_\mathrm{proj}/\lambda$ (see Table~\ref{tabobs}). It can be described
by
\begin{equation}
  |V| = |(1 - f_* - f_\mathrm{halo})V_\mathrm{ring} + f_*V_* +
  f_\mathrm{halo}V_\mathrm{halo}| \;,
\end{equation}
where $V_*=1$ is the visibility of the star, $V_\mathrm{ring}$ is the
visibility of the ring-shaped disk, $V_\mathrm{halo}$ is the
visibility of the halo, $f_\mathrm{halo}$ is the flux contribution of
the halo, and $1 - f_* - f_\mathrm{halo}$ is the flux contribution of
the ring-like disk. The wavelength-dependent relative flux
contribution $f_*$ of the stellar component to the total emission was
derived from the SED (Fig.~\ref{figsed}). The model parameters are the
inner ring radius $R_\mathrm{in}$ (semi-major axis in the elongated
case) of the extended ring, the inclination $i$, the position angle
$\vartheta$ of the semi-major axis of the elongated ring, and
$f_\mathrm{halo}$. An overview of our fit models, their parameters,
and the resulting $\chi^2_\mathrm{red}$ are presented in
Table~\ref{tabgeo}. We obtain inner radii between 0.78~AU
($\sim$$0.5$~mas) and 1.55~AU ($\sim$$1.0$~mas) and inclination angles
in the range of $\sim$$56\degr$ to $\sim$$70\degr$. The large error
bars of the NIR visibilities do not allow us to make a definite
decision on the elongation and the orientation of the disk as models
without elongation fit the data as well (see Table~\ref{tabgeo}).

\begin{table*}[htb]
  \caption{Overview of the geometric models. }
  \label{tabgeo}   
  \centering                       
  \begin{tabular}{lrrrrrr}  
    \hline\hline
    Model                            & $R_\mathrm{in}$ & $R_\mathrm{min}$ & $i$             & $\vartheta$       & $f_\mathrm{halo}$ & $\chi^2_\mathrm{red}$ \\
                                     & [AU]            & [AU]            & \degr           & \degr            &                  &         \\ \hline
    Ring                             & $1.08 \pm 0.38$ & $1.08 \pm 0.38$ & ...             & ...              & ...              & $2.69$ \\
    Ring + extended halo  & $0.78 \pm 0.31$ & $0.78 \pm 0.31$ & ...             & ...              & $0.09 \pm 0.01$  & $0.39$ \\
    Elongated ring                   & $1.55 \pm 0.53$ & $0.53 \pm 0.20$ & $70.0 \pm 0.01$ & $155.7 \pm 2.1$  & ...              & $1.88$ \\
    Elong. ring + extended halo & $1.07 \pm 0.43$ & $0.60 \pm 0.32$ & $55.8 \pm 0.09$ & $162.2 \pm 5.5$  & $0.09 \pm 0.01$  & $0.74$ \\
    \hline
  \end{tabular}
  \tablefoot{All models consist of the above ring-like disks, halos,
    and an unresolved stellar source. The distance uncertainty
    (Table~\ref{tabpro}) is taken into account in the above
    uncertainties for the radii. The radii $R_\mathrm{in}$ and
    $R_\mathrm{min}$ are the semi-major and the semi-minor axis of the
    ring models, respectively. The variable $\vartheta$ is the
    position angle of the semi-major axis.}
\end{table*}

%.....................................................................................
\subsection{Temperature-gradient model} \label{kaptgm}
To interpret our data, we model our visibility measurements and the
literature SED of HD~85567 with a temperature-gradient model. This
kind of modeling has already been used for modeling AMBER data from
young stars of different types by, for example, \citet{2005mallac},
\citet{2010krahof,2012kracal}, \citet{2012krekra}, \citet{2012chekre},
and \citet{2014vurkre}.

The temperature distribution $T(r)$ of a circumstellar disk is assumed
to be: $T(r)=T_\mathrm{in} \cdot \left(r/r_\mathrm{in}\right)^{-q}$,
where $r$ is the radius, $q$ is the temperature power-law index, and
$T_\mathrm{in}$ the temperature at the inner disk radius
$r_\mathrm{in}$. The disk is modeled as the sum of a large number of
narrow rings; each has a temperature $T(r)$. We integrate the
blackbody spectra of the rings and the intensity distribution of these
rings to obtain the total energy distribution and the visibilities. We
allow the model disk to be inclined, thus introducing the inclination
$i$ ($i=0\degr$ if the disk is face-on) and the position angle
$\vartheta$ (measured east of north) of the semi-major axis of the
model disk as additional parameters.

Our model includes a stellar component with the parameters from
Table~\ref{tabpro} (distance, $T_*$, $R_*$, and $L_*$) and a
surrounding circumstellar temperature-gradient disk extending to an
outer radius of $r_\mathrm{out}=r_\mathrm{in}+\Delta r$. Each computed
model contains the visibilities for all baseline-position angle
combinations in our measurements. We calculated the models for all
combinations of the parameter values as described in
Table~\ref{tabpar}, where each parameter was varied in 4 to 40 steps
within the described scan ranges. This method required the calculation
of approximately 170~million models.

\begin{table}[t]
 \centering  
  \caption{Scanned parameter space and best-fit temperature-gradient
    model described in Sect.~\ref{kaptgm}.}
  \label{tabpar}                     
  \begin{tabular}{llll} 
    \hline \hline
    Parameter                  & Scan range          & $N_\mathrm{val}$ & Best-fit value \\ 
    \hline
    $r_\mathrm{in}$ [AU]        & 0.04 -- 10          & 40              & $0.67^{+0.51}_{-0.15}$ \\
    $\Delta r$ [AU]            & 0.01 -- 50          & 30              & $24.8^{+20.5}_{-8.8}$  \\
    $T_\mathrm{in}$ [K]         & 200 -- 8000         & 40              & $2200^{+752}_{-341}$   \\
    $q$                        & 0.2, 0.5, 0.75, 1.0 & 4               & $0.75$                \\
    $i$ [\degr]                & 0 -- 90             & 30              & $52.6^{+14.7}_{-11.0}$  \\
    $\vartheta$ [\degr]        & 0 -- 170            & 30              & $121.4^{30.6+}_{-46.8}$ \\ 
    \hline
  \end{tabular}
  \tablefoot{$N_\mathrm{val}$ is the number of parameter values
    used. The spacing between the single parameter values is linear or
    as indicated, except for $r_\mathrm{in}$, where the spacing is
    logarithmic. For example, the parameter values for $r_\mathrm{in}$
    are 0.10~AU, 0.11~AU, 0.13~AU, ..., 10~AU.}
\end{table}

To find the best-fit temperature-gradient model, we calculated the
$\chi^2_\mathrm{red}$ (using
$\chi^2=\chi^2_\mathrm{SED}+\chi^2_\mathrm{Vis}$) of each model and
obtained a best-fit model with $\chi^2_\mathrm{red} \sim 1.0$ (see
Fig.~\ref{figmap} in the Appendix). The parameters of the best-fit
model are listed in Table~\ref{tabpar} and will be discussed in the
following section. The uncertainties are $3\sigma$ errors. The model
curves are shown in Figs.~\ref{figvis} and \ref{figsed}.

%__________________________________________________________________
\section{Discussion} \label{kapdis}

From our geometric modeling, we obtained different ring-fit radii for
different star-disk and star-disk-halo models (see
Table~\ref{tabgeo}). We achieve significantly better
$\chi^2_\mathrm{red}$ if the model has a halo component.  The
$\chi^2_\mathrm{red}$ closest to one was obtained for the geometric
ring model of an elongated ring and a halo. The inner ring radius of
this model is $R_\mathrm{in}\sim 1.1 \pm 0.4$~AU. \citet{2013whewei}
estimated the ring radius with a symmetric star-ring model including
resolved background emission to $0.8 \pm 0.3$~AU. This value is
consistent with most of our derived ring-fit radii
(Table~\ref{tabgeo}). 
The high visibility values and small closures phases (which are
approximately zero; see Fig.~\ref{figcp}) of our observations and of
the observations reported by \citet{2013whewei} do not allow us to
detect any binary signature. Binarity is believed to be a key property
in B[e] stars \citep{2007mir}.

\begin{figure}[tb]
  \includegraphics[angle=-90,width=0.5\textwidth]{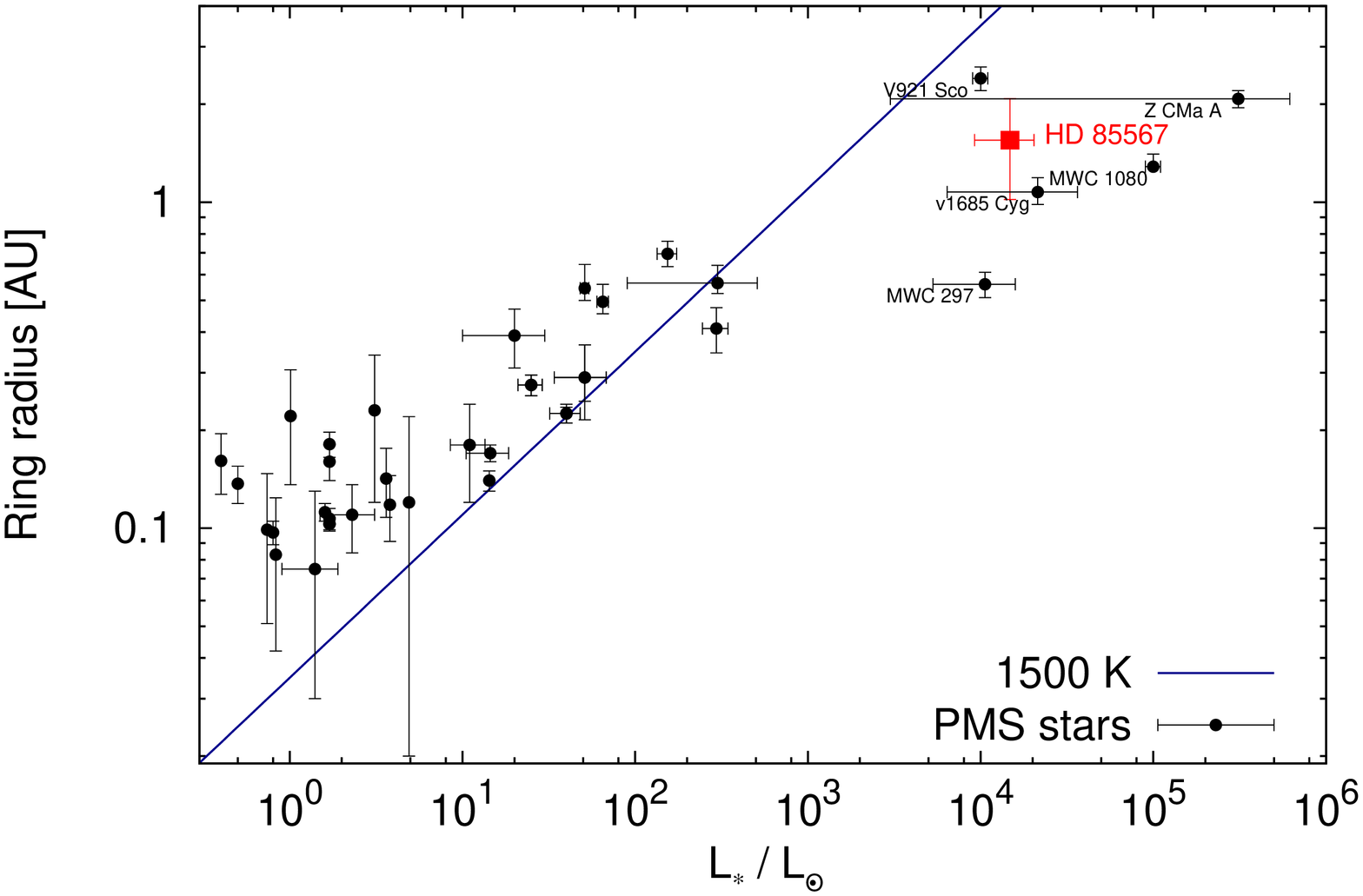}
  \caption{The position of HD~85567 (red square) in the
    size-luminosity diagram of T~Tauri stars, Herbig Ae, and Herbig Be
    stars. The data are taken from the literature
    \citep{2004eislan,2005monmil,2008krahof,2008pinmen,2011weigri,2012chekre,2012krekra,2012vurkre,2012wanwei,2013krewei,2014vurkre}. The
    solid line shows the theoretical dependence of the inner ring
    radius on the luminosity for an optically thin disk hole and a
    dust sublimation temperature of 1500~K \citep{2005monmil}.}
  \label{figslr}
\end{figure}

Our geometric ring-fit radii ($\sim$0.8--1.6~AU) are approximately
3--5 times smaller than the radii predicted by the size-luminosity
relation ($\sim$4.2~AU; Fig.~\ref{figslr}). The predicted radii of the
size-luminosity relation \citep{2005monmil} are based on the
assumption that the dust sublimates at a temperature of 1500~K and the
inner cavity is dust-free and optically thin. These assumptions are
approximately valid for HAe stars. However, several interferometric
observations suggest that some of the more massive HBe stars harbor a
gaseous, optically thick disk inside the dust disk. This gas disk can
partially shield the stellar radiation, thereby letting dust survive
closer to the star
\citep{2002monmil,2005monmil,2007milmal,2011weigri}.

Our best-fit temperature-gradient model (Table~\ref{tabpar},
Figs.~\ref{figvis}, \ref{figsed}) consists of the stellar point source
and an extended disk with an inner radius of
$r_\mathrm{in}=0.67^{+0.51}_{-0.21}$~AU. This inner radius is
approximately 1.6 to 2.3 times smaller than the elongated ring-fit
radii derived with geometric modeling (Table~\ref{tabgeo}) but also
corresponds to a higher temperature. We derived a disk inclination of
$i=52.6^{+14.7}_{-11.0}${\degr} and a position angle of the disk
semi-major axis of $\vartheta=121.4^{30.6+}_{-46.8}${\degr}, which are
similar to the values derived with geometric modeling, but not very
well constrained due to the large error bars of the NIR visibilities
(cf.  Fig.~\ref{figmap} in the Appendix). The derived inner
temperature $T_\mathrm{in}=2200^{+752}_{-341}$~K is too hot for the
standard dust composition consisting mainly of silicates, but the
existence of refractory dust grains \citep[e.g., iron, graphite,
  corundum;][]{2010bennat} may explain this high temperature. Gas
emission inside the dust disk can also contribute to the {\textit
  K}-band visibility and make the average NIR radius appear smaller
(e.g., \citealt{2008krapre}). A size estimate based on the accretion
rate supports the presence of an optically thick inner disk
\citep{2013whewei}. By modeling the CO bandhead emission, these
authors found that a compact (inner radius approximately 0.2--1~AU),
optically thick gas disk can reproduce their measurements. This agrees
with the hypothesis of shielded stellar radiation, which leads to
small dust sublimation radii, which agrees with our observations.

%__________________________________________________________________
\section{Conclusion}
 We used geometric and temperature-gradient models to interpret our
 VLTI/AMBER data and the SED of the Herbig B[e] star HD~85567. We
 derived geometric ring models with inner radii of 0.8--1.6~AU, which
 are approximately 3--5 times smaller than that predicted by the
 size-luminosity relation. Using temperature-gradient modeling, we
 found a hot ($T_\mathrm{in}\sim 2200$~K) inner disk rim with a small
 ($r_\mathrm{in}\sim 0.67$~AU) inner radius. The undersized inner disk
 radius obtained with both geometric and temperature-gradient modeling
 agrees with measurements of several other HBe stars and with previous
 measurements of this object. The small inner radius provides further
 support to the existence of an optically thick gaseous inner disk
 shielding the stellar radiation in Herbig Be stars.

%__________________________________________________________________
\begin{acknowledgements}
  This research has made use of the NASA/ IPAC Infrared Science
  Archive, which is operated by the Jet Propulsion Laboratory,
  California Institute of Technology, under contract with the National
  Aeronautics and Space Administration. We thank our ESO colleagues at
  the Paranal observatory for their excellent collaboration.
\end{acknowledgements}

\bibliographystyle{aa} \bibliography{BIBO}

\begin{thebibliography}{36}
\expandafter\ifx\csname natexlab\endcsname\relax\def\natexlab#1{#1}\fi

\bibitem[{{Baines} {et~al.}(2006){Baines}, {Oudmaijer}, {Porter}, \&
  {Pozzo}}]{2006baioud}
{Baines}, D., {Oudmaijer}, R.~D., {Porter}, J.~M., \& {Pozzo}, M. 2006, \mnras,
  367, 737

\bibitem[{{Benisty} {et~al.}(2010){Benisty}, {Natta}, {Isella}, {Berger},
  {Massi}, {Le Bouquin}, {M{\'e}rand}, {Duvert}, {Kraus}, {Malbet}, {Olofsson},
  {Robbe-Dubois}, {Testi}, {Vannier}, \& {Weigelt}}]{2010bennat}
{Benisty}, M., {Natta}, A., {Isella}, A., {et~al.} 2010, \aap, 511, A74

\bibitem[{{Chelli} {et~al.}(2009){Chelli}, {Utrera}, \& {Duvert}}]{2009cheutr}
{Chelli}, A., {Utrera}, O.~H., \& {Duvert}, G. 2009, \aap, 502, 705

\bibitem[{{Chen} {et~al.}(2012){Chen}, {Kreplin}, {Wang}, {Weigelt}, {Hofmann},
  {Kraus}, {Schertl}, {Lagarde}, {Natta}, {Petrov}, {Robbe-Dubois}, \&
  {Tatulli}}]{2012chekre}
{Chen}, L., {Kreplin}, A., {Wang}, Y., {et~al.} 2012, \aap, 541, A104

\bibitem[{{Dullemond} {et~al.}(2001){Dullemond}, {Dominik}, \&
  {Natta}}]{2001duldom}
{Dullemond}, C.~P., {Dominik}, C., \& {Natta}, A. 2001, \apj, 560, 957

\bibitem[{{Dullemond} \& {Monnier}(2010)}]{2010dulmon}
{Dullemond}, C.~P. \& {Monnier}, J.~D. 2010, \araa, 48, 205

\bibitem[{{Eisner} {et~al.}(2003){Eisner}, {Lane}, {Akeson}, {Hillenbrand}, \&
  {Sargent}}]{2003eislan}
{Eisner}, J.~A., {Lane}, B.~F., {Akeson}, R.~L., {Hillenbrand}, L.~A., \&
  {Sargent}, A.~I. 2003, \apj, 588, 360

\bibitem[{{Eisner} {et~al.}(2004){Eisner}, {Lane}, {Hillenbrand}, {Akeson}, \&
  {Sargent}}]{2004eislan}
{Eisner}, J.~A., {Lane}, B.~F., {Hillenbrand}, L.~A., {Akeson}, R.~L., \&
  {Sargent}, A.~I. 2004, \apj, 613, 1049

\bibitem[{{Ilee}(2013)}]{2013ile}
{Ilee}, J.~D. 2013, PhD thesis, University of Leeds

\bibitem[{{Kraus} {et~al.}(2012){Kraus}, {Calvet}, {Hartmann}, {Hofmann},
  {Kreplin}, {Monnier}, \& {Weigelt}}]{2012kracal}
{Kraus}, S., {Calvet}, N., {Hartmann}, L., {et~al.} 2012, \apj, 752, 11

\bibitem[{{Kraus} {et~al.}(2008{\natexlab{a}}){Kraus}, {Hofmann}, {Benisty},
  {Berger}, {Chesneau}, {Isella}, {Malbet}, {Meilland}, {Nardetto}, {Natta},
  {Preibisch}, {Schertl}, {Smith}, {Stee}, {Tatulli}, {Testi}, \&
  {Weigelt}}]{2008krahof}
{Kraus}, S., {Hofmann}, K.-H., {Benisty}, M., {et~al.} 2008{\natexlab{a}},
  \aap, 489, 1157

\bibitem[{{Kraus} {et~al.}(2010){Kraus}, {Hofmann}, {Menten}, {Schertl},
  {Weigelt}, {Wyrowski}, {Meilland}, {Perraut}, {Petrov}, {Robbe-Dubois},
  {Schilke}, \& {Testi}}]{2010krahof}
{Kraus}, S., {Hofmann}, K.-H., {Menten}, K.~M., {et~al.} 2010, \nat, 466, 339

\bibitem[{{Kraus} {et~al.}(2008{\natexlab{b}}){Kraus}, {Preibisch}, \&
  {Ohnaka}}]{2008krapre}
{Kraus}, S., {Preibisch}, T., \& {Ohnaka}, K. 2008{\natexlab{b}}, \apj, 676,
  490

\bibitem[{{Kreplin} {et~al.}(2012){Kreplin}, {Kraus}, {Hofmann}, {Schertl},
  {Weigelt}, \& {Driebe}}]{2012krekra}
{Kreplin}, A., {Kraus}, S., {Hofmann}, K.-H., {et~al.} 2012, \aap, 537, A103

\bibitem[{{Kreplin} {et~al.}(2013){Kreplin}, {Weigelt}, {Kraus}, {Grinin},
  {Hofmann}, {Kishimoto}, {Schertl}, {Tambovtseva}, {Clausse}, {Massi},
  {Perraut}, \& {Stee}}]{2013krewei}
{Kreplin}, A., {Weigelt}, G., {Kraus}, S., {et~al.} 2013, \aap, 551, A21

\bibitem[{{Lafrasse} {et~al.}(2010){Lafrasse}, {Mella}, {Bonneau}, {Duvert},
  {Delfosse}, \& {Chelli}}]{2010lafmel}
{Lafrasse}, S., {Mella}, G., {Bonneau}, D., {et~al.} 2010, VizieR Online Data
  Catalog, 2300, 0

\bibitem[{{Lamers} {et~al.}(1998){Lamers}, {Zickgraf}, {de Winter}, {Houziaux},
  \& {Zorec}}]{1998lamzic}
{Lamers}, H.~J.~G.~L.~M., {Zickgraf}, F.-J., {de Winter}, D., {Houziaux}, L.,
  \& {Zorec}, J. 1998, \aap, 340, 117

\bibitem[{{Malbet} {et~al.}(2005){Malbet}, {Lachaume}, {Berger}, {Colavita},
  {di Folco}, {Eisner}, {Lane}, {Millan-Gabet}, {S{\'e}gransan}, \&
  {Traub}}]{2005mallac}
{Malbet}, F., {Lachaume}, R., {Berger}, J.-P., {et~al.} 2005, \aap, 437, 627

\bibitem[{{Malfait} {et~al.}(1998){Malfait}, {Bogaert}, \&
  {Waelkens}}]{1998malbog}
{Malfait}, K., {Bogaert}, E., \& {Waelkens}, C. 1998, \aap, 331, 211

\bibitem[{{Millan-Gabet} {et~al.}(2007){Millan-Gabet}, {Malbet}, {Akeson},
  {Leinert}, {Monnier}, \& {Waters}}]{2007milmal}
{Millan-Gabet}, R., {Malbet}, F., {Akeson}, R., {et~al.} 2007, Protostars and
  Planets V, 539

\bibitem[{{Millan-Gabet} {et~al.}(2001){Millan-Gabet}, {Schloerb}, \&
  {Traub}}]{2001milsch}
{Millan-Gabet}, R., {Schloerb}, F.~P., \& {Traub}, W.~A. 2001, \apj, 546, 358

\bibitem[{{Miroshnichenko}(2007)}]{2007mir}
{Miroshnichenko}, A.~S. 2007, \apj, 667, 497

\bibitem[{{Miroshnichenko} {et~al.}(2001){Miroshnichenko}, {Levato},
  {Bjorkman}, \& {Grosso}}]{2001mirlev}
{Miroshnichenko}, A.~S., {Levato}, H., {Bjorkman}, K.~S., \& {Grosso}, M. 2001,
  \aap, 371, 600

\bibitem[{{Monnier} \& {Millan-Gabet}(2002)}]{2002monmil}
{Monnier}, J.~D. \& {Millan-Gabet}, R. 2002, \apj, 579, 694

\bibitem[{{Monnier} {et~al.}(2005){Monnier}, {Millan-Gabet}, {Billmeier},
  {Akeson}, {Wallace}, {Berger}, {Calvet}, {D'Alessio}, {Danchi}, {Hartmann},
  {Hillenbrand}, {Kuchner}, {Rajagopal}, {Traub}, {Tuthill}, {Boden}, {Booth},
  {Colavita}, {Gathright}, {Hrynevych}, {Le Mignant}, {Ligon}, {Neyman},
  {Swain}, {Thompson}, {Vasisht}, {Wizinowich}, {Beichman}, {Beletic},
  {Creech-Eakman}, {Koresko}, {Sargent}, {Shao}, \& {van Belle}}]{2005monmil}
{Monnier}, J.~D., {Millan-Gabet}, R., {Billmeier}, R., {et~al.} 2005, \apj,
  624, 832

\bibitem[{{Natta} {et~al.}(2001){Natta}, {Prusti}, {Neri}, {Wooden}, {Grinin},
  \& {Mannings}}]{2001natpru}
{Natta}, A., {Prusti}, T., {Neri}, R., {et~al.} 2001, \aap, 371, 186

\bibitem[{{Pasinetti Fracassini} {et~al.}(2001){Pasinetti Fracassini},
  {Pastori}, {Covino}, \& {Pozzi}}]{2001paspas}
{Pasinetti Fracassini}, L.~E., {Pastori}, L., {Covino}, S., \& {Pozzi}, A.
  2001, \aap, 367, 521

\bibitem[{{Petrov} {et~al.}(2007){Petrov}, {Malbet}, {Weigelt}, {Antonelli},
  {Beckmann}, {Bresson}, {Chelli}, {Dugu{\'e}}, {Duvert}, {Gennari},
  {Gl{\"u}ck}, {Kern}, {Lagarde}, {Le Coarer}, {Lisi}, {Millour}, {Perraut},
  {Puget}, {Rantakyr{\"o}}, {Robbe-Dubois}, {Roussel}, {Salinari}, {Tatulli},
  {Zins}, {Accardo}, {Acke}, {Agabi}, {Altariba}, {Arezki}, {Aristidi},
  {Baffa}, {Behrend}, {Bl{\"o}cker}, {Bonhomme}, {Busoni}, {Cassaing},
  {Clausse}, {Colin}, {Connot}, {Delboulb{\'e}}, {Domiciano de Souza},
  {Driebe}, {Feautrier}, {Ferruzzi}, {Forveille}, {Fossat}, {Foy},
  {Fraix-Burnet}, {Gallardo}, {Giani}, {Gil}, {Glentzlin}, {Heiden},
  {Heininger}, {Hernandez Utrera}, {Hofmann}, {Kamm}, {Kiekebusch}, {Kraus},
  {Le Contel}, {Le Contel}, {Lesourd}, {Lopez}, {Lopez}, {Magnard}, {Marconi},
  {Mars}, {Martinot-Lagarde}, {Mathias}, {M{\`e}ge}, {Monin}, {Mouillet},
  {Mourard}, {Nussbaum}, {Ohnaka}, {Pacheco}, {Perrier}, {Rabbia}, {Rebattu},
  {Reynaud}, {Richichi}, {Robini}, {Sacchettini}, {Schertl}, {Sch{\"o}ller},
  {Solscheid}, {Spang}, {Stee}, {Stefanini}, {Tallon}, {Tallon-Bosc}, {Tasso},
  {Testi}, {Vakili}, {von der L{\"u}he}, {Valtier}, {Vannier}, \&
  {Ventura}}]{2007petmal}
{Petrov}, R.~G., {Malbet}, F., {Weigelt}, G., {et~al.} 2007, \aap, 464, 1

\bibitem[{{Pinte} {et~al.}(2008){Pinte}, {M{\'e}nard}, {Berger}, {Benisty}, \&
  {Malbet}}]{2008pinmen}
{Pinte}, C., {M{\'e}nard}, F., {Berger}, J.~P., {Benisty}, M., \& {Malbet}, F.
  2008, \apjl, 673, L63

\bibitem[{{Tatulli} {et~al.}(2007){Tatulli}, {Millour}, {Chelli}, {Duvert},
  {Acke}, {Hernandez Utrera}, {Hofmann}, {Kraus}, {Malbet}, {M{\`e}ge},
  {Petrov}, {Vannier}, {Zins}, {Antonelli}, {Beckmann}, {Bresson}, {Dugu{\'e}},
  {Gennari}, {Gl{\"u}ck}, {Kern}, {Lagarde}, {Le Coarer}, {Lisi}, {Perraut},
  {Puget}, {Rantakyr{\"o}}, {Robbe-Dubois}, {Roussel}, {Weigelt}, {Accardo},
  {Agabi}, {Altariba}, {Arezki}, {Aristidi}, {Baffa}, {Behrend}, {Bl{\"o}cker},
  {Bonhomme}, {Busoni}, {Cassaing}, {Clausse}, {Colin}, {Connot},
  {Delboulb{\'e}}, {Domiciano de Souza}, {Driebe}, {Feautrier}, {Ferruzzi},
  {Forveille}, {Fossat}, {Foy}, {Fraix-Burnet}, {Gallardo}, {Giani}, {Gil},
  {Glentzlin}, {Heiden}, {Heininger}, {Kamm}, {Kiekebusch}, {Le Contel}, {Le
  Contel}, {Lesourd}, {Lopez}, {Lopez}, {Magnard}, {Marconi}, {Mars},
  {Martinot-Lagarde}, {Mathias}, {Monin}, {Mouillet}, {Mourard}, {Nussbaum},
  {Ohnaka}, {Pacheco}, {Perrier}, {Rabbia}, {Rebattu}, {Reynaud}, {Richichi},
  {Robini}, {Sacchettini}, {Schertl}, {Sch{\"o}ller}, {Solscheid}, {Spang},
  {Stee}, {Stefanini}, {Tallon}, {Tallon-Bosc}, {Tasso}, {Testi}, {Vakili},
  {von der L{\"u}he}, {Valtier}, \& {Ventura}}]{2007tatmil}
{Tatulli}, E., {Millour}, F., {Chelli}, A., {et~al.} 2007, \aap, 464, 29

\bibitem[{{Verhoeff} {et~al.}(2012){Verhoeff}, {Waters}, {van den Ancker},
  {Min}, {Stap}, {Pantin}, {van Boekel}, {Acke}, {Tielens}, \& {de
  Koter}}]{2012verwat}
{Verhoeff}, A.~P., {Waters}, L.~B.~F.~M., {van den Ancker}, M.~E., {et~al.}
  2012, \aap, 538, A101

\bibitem[{{Vural} {et~al.}(2014){Vural}, {Kreplin}, {Kishimoto}, {Weigelt},
  {Hofmann}, {Kraus}, {Schertl}, {Dugu{\'e}}, {Duvert}, {Lagarde}, \&
  {Massi}}]{2014vurkre}
{Vural}, J., {Kreplin}, A., {Kishimoto}, M., {et~al.} 2014, \aap, 564, A118

\bibitem[{{Vural} {et~al.}(2012){Vural}, {Kreplin}, {Kraus}, {Weigelt},
  {Driebe}, {Benisty}, {Dugu{\'e}}, {Massi}, {Monin}, \&
  {Vannier}}]{2012vurkre}
{Vural}, J., {Kreplin}, A., {Kraus}, S., {et~al.} 2012, \aap, 543, A162

\bibitem[{{Wang} {et~al.}(2012){Wang}, {Weigelt}, {Kreplin}, {Hofmann},
  {Kraus}, {Miroshnichenko}, {Schertl}, {Chelli}, {Domiciano de Souza},
  {Massi}, \& {Robbe-Dubois}}]{2012wanwei}
{Wang}, Y., {Weigelt}, G., {Kreplin}, A., {et~al.} 2012, \aap, 545, L10

\bibitem[{{Weigelt} {et~al.}(2011){Weigelt}, {Grinin}, {Groh}, {Hofmann},
  {Kraus}, {Miroshnichenko}, {Schertl}, {Tambovtseva}, {Benisty}, {Driebe},
  {Lagarde}, {Malbet}, {Meilland}, {Petrov}, \& {Tatulli}}]{2011weigri}
{Weigelt}, G., {Grinin}, V.~P., {Groh}, J.~H., {et~al.} 2011, \aap, 527, A103

\bibitem[{{Wheelwright} {et~al.}(2013){Wheelwright}, {Weigelt}, {Caratti o
  Garatti}, \& {Garcia Lopez}}]{2013whewei}
{Wheelwright}, H.~E., {Weigelt}, G., {Caratti o Garatti}, A., \& {Garcia
  Lopez}, R. 2013, \aap, 558, A116

\end{thebibliography}

\newpage
\begin{appendix}
  \section{$\chi^2$-maps}
  \begin{figure}[hb]
    \includegraphics[angle=-90, width=0.5\textwidth]{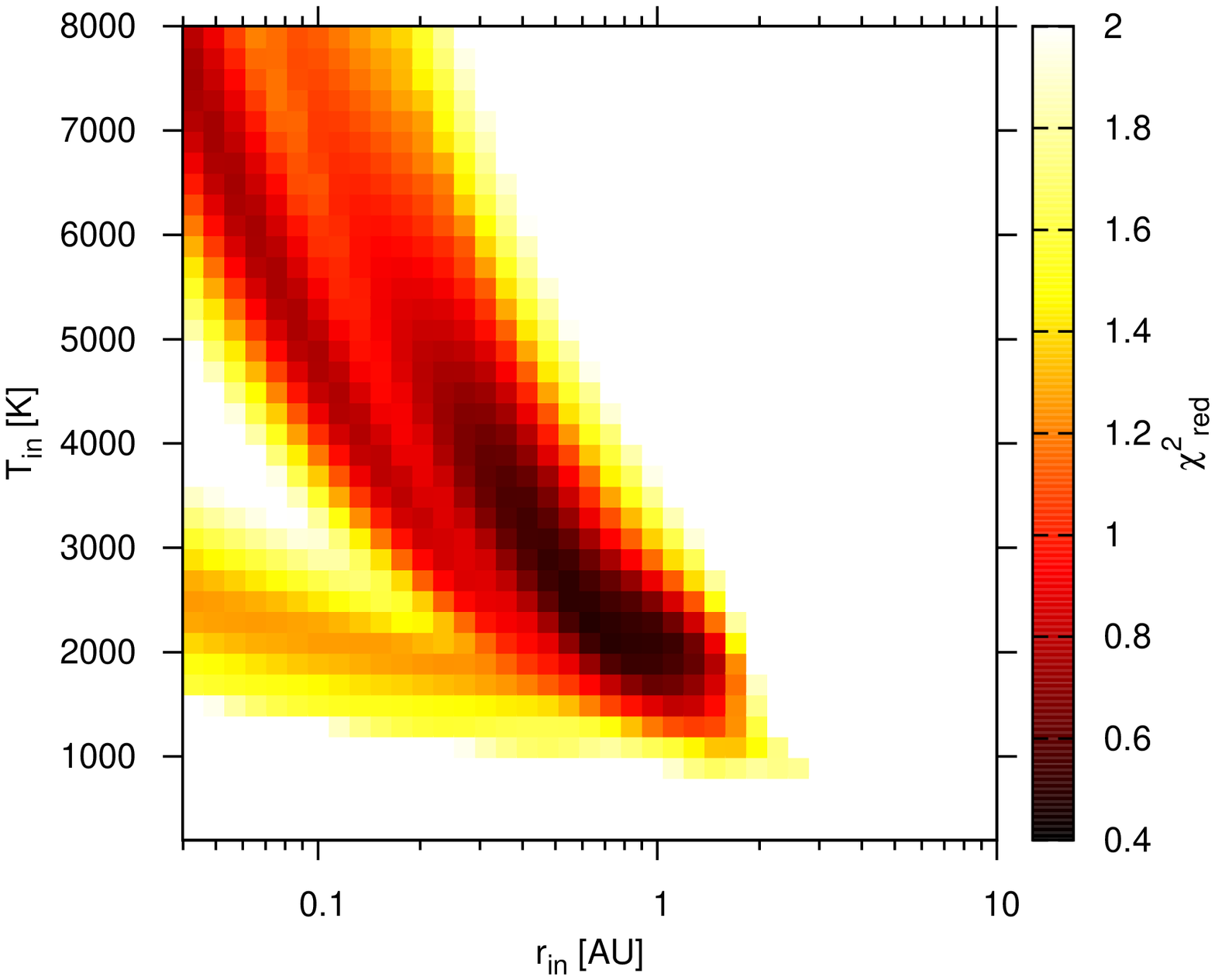}
    \includegraphics[angle=-90, width=0.5\textwidth]{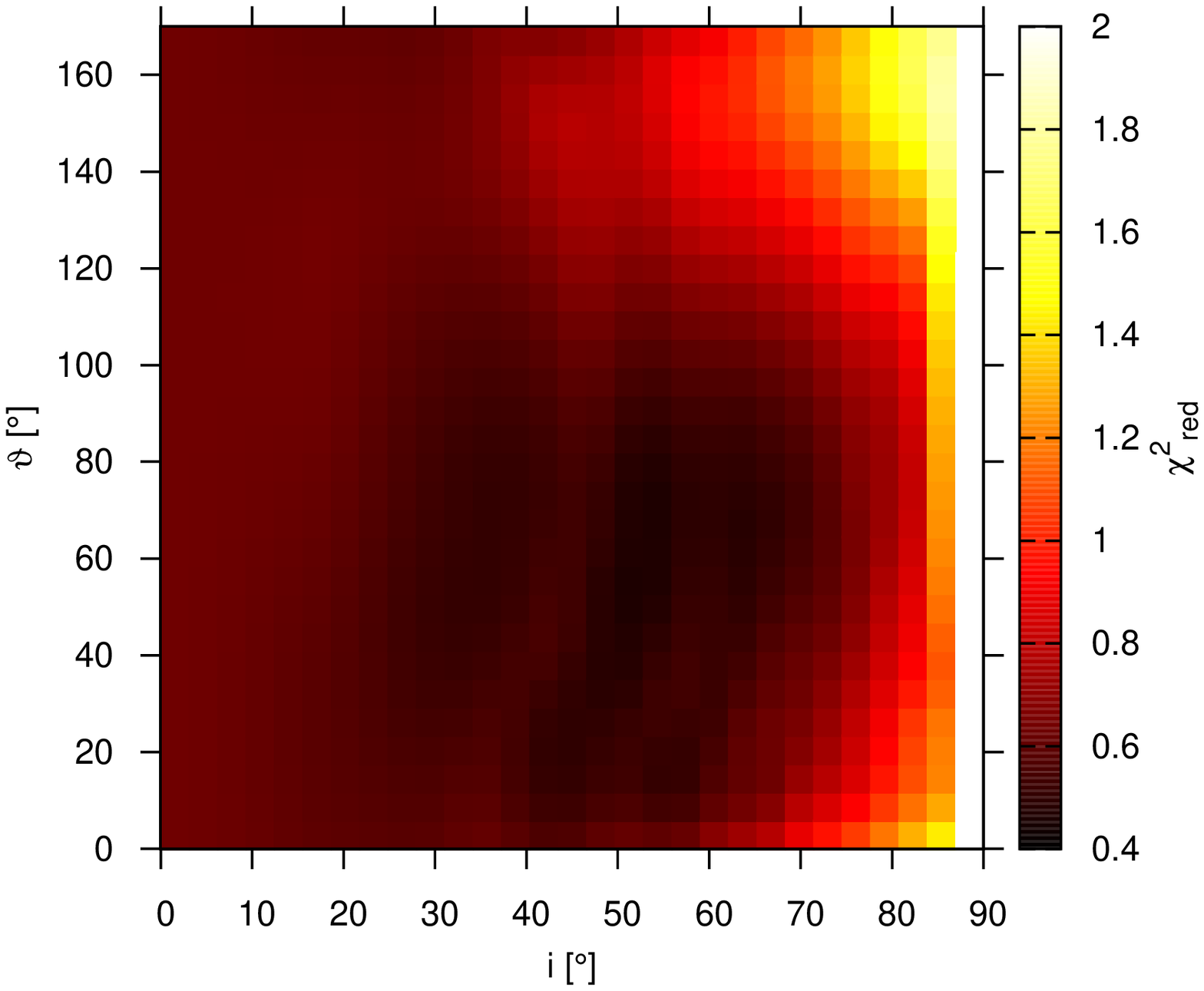}
    \caption{Two-dimensional $\chi^2_\mathrm{red}$ distributions of
      selected parameters of the temperature-gradient model.}
    \label{figmap}
  \end{figure}
\end{appendix}

\end{document}